\documentclass[twocolumn]{aastex6}

\usepackage{hyperref}

\newcommand{\grb}{GRB~151006A}
\newcommand{\swift}{\textit{Swift}}
\newcommand{\fermi}{\textit{Fermi}}
\newcommand{\sw}[1]{\texttt{#1}}
\newcommand{\edi}[1]{#1}

\slugcomment{\apj\, submitted}

\shorttitle{CZT Imager observations of  GRB~151006A}
\shortauthors{Rao et al.}

\begin{document}

\title{AstroSat CZT Imager observations of  GRB~151006A: timing, spectroscopy, and polarisation study}

\author{A. R. Rao\altaffilmark{1},
Vikas Chand\altaffilmark{1}, 
M.K. Hingar\altaffilmark{1}, 
S. Iyyani\altaffilmark{1},
Rakesh Khanna\altaffilmark{1},
A.P.K. Kutty\altaffilmark{1},
J.P. Malkar\altaffilmark{1}, 
D. Paul\altaffilmark{1},
V. B. Bhalerao\altaffilmark{2}, 
D. Bhattacharya\altaffilmark{2},
G. C. Dewangan\altaffilmark{2},
Pramod Pawar\altaffilmark{2,3},
A. M. Vibhute\altaffilmark{2},
T. Chattopadhyay\altaffilmark{4}, 
N.P.S. Mithun\altaffilmark{4},
S.V. Vadawale\altaffilmark{4}, 
N. Vagshette\altaffilmark{4}, 
R. Basak\altaffilmark{5},
P. Pradeep\altaffilmark{6}, 
Essy Samuel\altaffilmark{6},
S. Sreekumar\altaffilmark{6}, 
P. Vinod\altaffilmark{6}, 
K.H. Navalgund\altaffilmark{7}, 
R. Pandiyan\altaffilmark{7},
K. S. Sarma\altaffilmark{7}, 
S. Seetha\altaffilmark{7}
K. Subbarao\altaffilmark{7}
} 

\altaffiltext{1}{Department of Astronomy and Astrophysics, Tata Institute of Fundamental Research, Homi Bhabha Road, Mumbai, India; \email{arrao@tifr.res.in}}

\altaffiltext{2}{Inter University Center for Astronomy \& Astrophysics, Pune, India}

\altaffiltext{3}{ S. R. T. M. University, Nanded, India}

\altaffiltext{4}{Physical Research Laboratory, Ahmedabad, India }
 
\altaffiltext{5}{Nicolaus Copernicus Astronomical Center, Polish Academy of Sciences, Warsaw, Poland}

\altaffiltext{6}{Vikram Sarabhai Space Centre, Thiruvananthapuram, India}

\altaffiltext{7}{ISRO Satellite Centre, Bengaluru, India}

\begin{abstract}
$AstroSat$ is a multi-wavelength satellite launched on 2015 September 28. The CZT Imager of $AstroSat$ on its very first day of operation detected a long duration gamma-ray burst (GRB) namely GRB 151006A.
Using the off-axis imaging and spectral response of the instrument,
we demonstrate that CZT Imager can localise this GRB correct to about a few degrees and it
can provide, in conjunction with \swift, spectral parameters similar to that obtained 
from \fermi/GBM. 
Hence CZT Imager would be a useful addition to the currently operating GRB instruments (\swift\ and \fermi). Specifically, we argue that the CZT Imager will be most useful for the short hard GRBs by providing localisation for those detected by \fermi\ and spectral information for those detected only by \swift.
We also provide preliminary results on 
a new exciting capability of this instrument: CZT Imager is able to identify Compton
scattered events thereby providing polarisation information for bright GRBs. 
GRB~151006A, in spite of being relatively faint,  \edi{shows hints of a polarisation signal at 100--300 keV (though at a low significance level). We point out that CZT Imager should provide significant time resolved polarisation measurements for GRBs that have fluence 3 times higher than that of GRB~151006A}.
We estimate that the number of such bright GRBs detectable by CZT Imager is 
5 -- 6 per year. CZT Imager can also act as a good hard X--ray monitoring 
device for possible electromagnetic counterparts of Gravitational Wave events.

\end{abstract}


\section{Introduction}

The past decade has seen a tremendous improvement in our understanding of gamma-ray bursts (GRBs), particularly after the launch of \swift\  and \fermi\ satellites \citep{SwiftREF, GehrelsSci2012}.

With its quick localisation ability, \swift\ could detect the afterglows of many GRBs and help   measure their redshifts \citep*{GehrelsARAA2009}.  The \fermi\ satellite, on the other hand, provided the widest ever spectral coverage of the prompt emission of GRBs from 8 keV to $\sim40$ MeV using the GBM instrument, and extending further up to $>$300 GeV with the LAT instrument for some GRBs~\citep{Meegan2009, Atwood2009}. 

\begin{figure*}[ht]
\begin{centering}
\plottwo{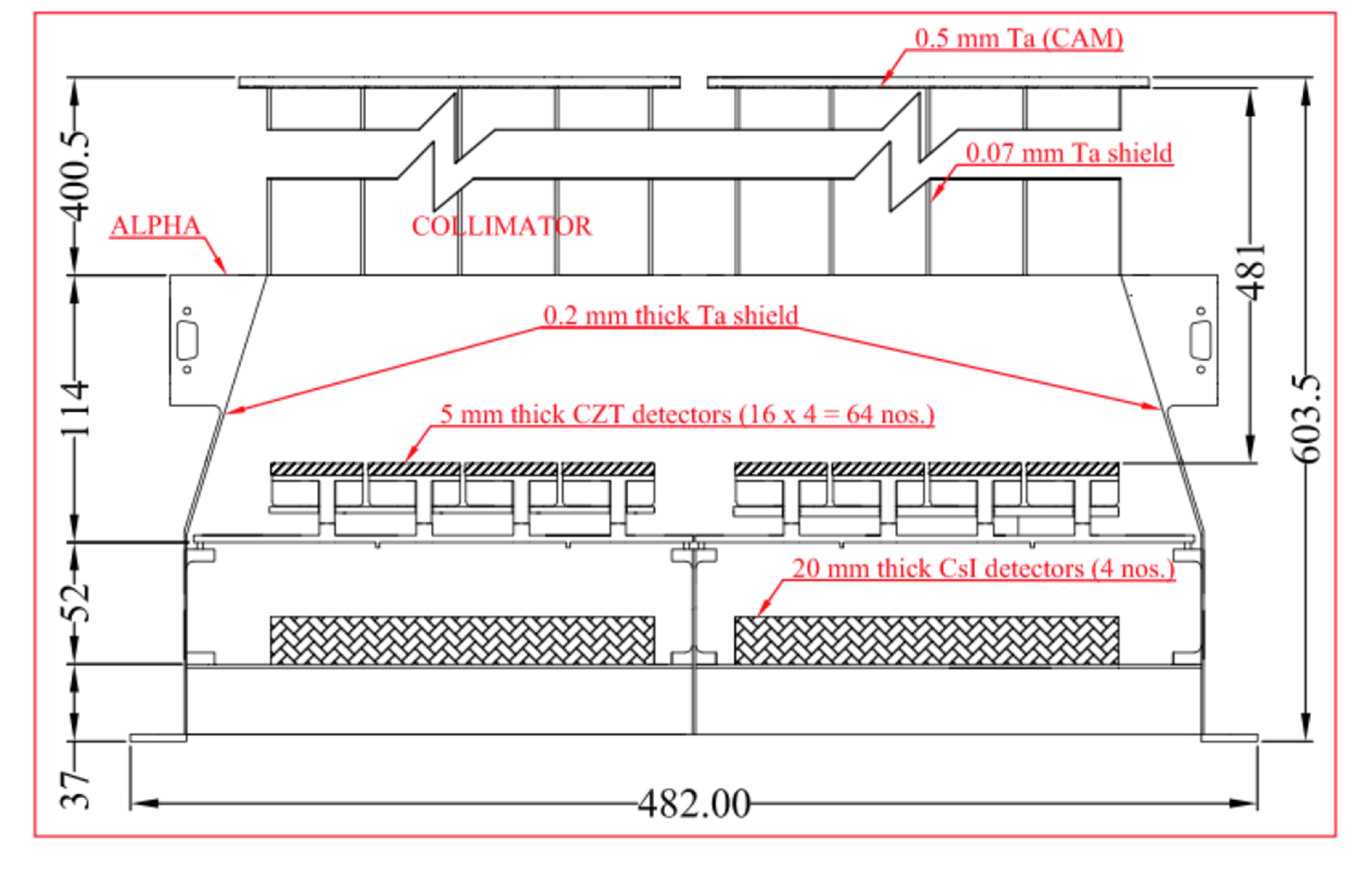}{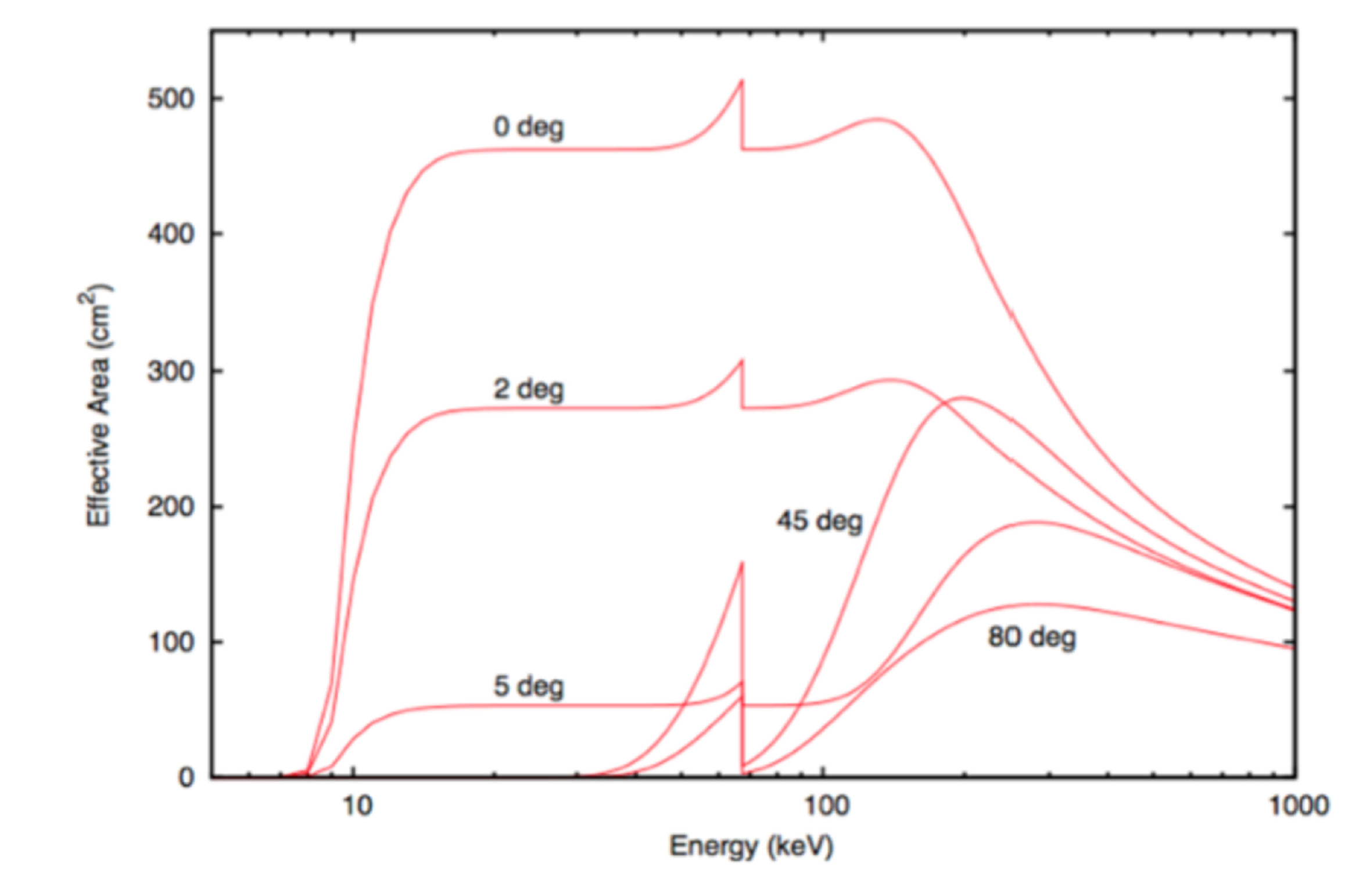}
\end{centering}
\caption{{\emph Left:} A schematic diagram of the CZT Imager instrument onboard the {\em AstroSat}~satellite. \edi{All dimensions are in mm}. A Tantalum coded aperture mask (CAM) at the top is backed by 400~mm Al/Ta collimators, which restrict the field of view to $4\degr.6\times 4\degr.6$. Four identical quadrants with $4\times4$ arrays of 5~mm thick CZT modules forms the focal plane, 481~mm below the CAM. The 2.46~mm pixels matched to the CAM pitch provide a native angular resolution of $\sim17\arcmin$ in the primary field of view. 20~mm thick CsI(Tl) scintillators mounted $\sim66$~mm below the CZT modules provide active anti--coincidence shielding, and also function as a wide--angle detector in the 100--500~keV range.
{\emph Right:}  Effective area of CZT Imager as a function of energy. The different lines are for sources at different angles from the CZTI viewing axis (marked in the figure). The sharp rise at $\sim57$~keV is due to the K--edge of Ta. }
\label{fig:CZTIimage}
\end{figure*}

A new addition to the suite of instruments studying GRBs is the hard X--ray imager Cadmium Zinc Telluride Imager (CZTI) on {\em AstroSat}, the Indian multi--wavelength observatory~\citep{Singh2014}. CZTI utilises a coded aperture mask and Cadmium Zinc Telluride detectors (Figure~\ref{fig:CZTIimage}, left) to image a 4.6\degr $\times$ 4.6\degr\ area of the sky in the 20--200~keV range \citep{Bhalerao2016}. Apart from this primary coded field of view, CZTI functions as an open detector at energies $>100$~keV, sensitive to almost the entire sky (Figure~\ref{fig:CZTIimage}, right). At these energies, CZTI also has X--ray polarisation capabilities \citep{Tanmoy14, Santosh15}. CsI (Tl) scintillators placed below the CZT modules for active anti--coincidence shielding (Figure~\ref{fig:CZTIimage}, right) also serve as all--sky high energy detectors in the 100--500~keV range. 

\swift\ and \fermi\ satellites showcase two different approaches to the study of GRBs. 
For quick and precise localisation, \swift\ uses a Coded Aperture Mask (CAM) and large area pixelated Cadmium Zinc Telluride (CZT) detectors with 4~mm$\times4$~mm pixels of 2~mm thickness \citep{BATREF}. At such thickness, CZT detectors have low spectral response at higher energies (above about 150~keV) and hence BAT by itself cannot precisely measure the peak energy of hard GRBs. In fact the number of short GRBs detected by \swift\ is much lower than that detected by other instruments, particularly due to the lack of response to high energy X--rays \citep{Band2006}. 
On the other hand, \fermi\ uses multiple open NaI crystal detectors \citep{Meegan2009} and localises the GRBs by comparing the relative counts in the different detectors. Hence the resultant  localisation accuracy is poor (several degrees) and the energy resolution of the detectors are rather modest \citep{Meegan2009}. Consequently, the prompt spectral studies are hampered and hence a good understanding of the radiation mechanism during the prompt phase is lacking. With its wide field of view, high energy coverage, and good spectral resolution, CZTI thus fills the gap between the capabilities \swift\ and \fermi.

On the very first day of operation, CZTI detected  \grb~\citep{Bhalerao2015}. \grb\ was first reported by {\em Swift} BAT~\citep{Kocevski2015} at $\alpha = 09^{\rm h}49^{\rm m}48^{\rm s}$, $\delta = +70\degr30\arcmin31\arcsec$. The prompt emission lasted for more than 300~s and, subsequently, detection by other X--ray and Gamma-ray missions were reported in a series of GCN circulars. \fermi\ triggered $\sim4$~s before {\em BAT} and reported a T$_{90}$ of $\sim84$~s \citep{Roberts15}. Due to the wide angle detection capabilities of  CZTI, {\grb} was registered in CZTI even though the incident  angle was as large as $\sim 60^{\circ}$ from its pointing direction \citep{Bhalerao2015}.  The presence of double Compton events are also seen in CZTI, which will help measure the polarisation of the GRB \citep{Santosh15}.  Coincidentally, \grb\ is also the first detected GRB in the CALorimetric Electron Telescope (CALET) aboard the International Space Station Gamma-Ray Burst Monitor detection (CGBM).  CGBM reported a double peak separated by 4 seconds \citep{Yoshida15}.

In this paper, we present the results of the CZTI observations of \grb. We demonstrate that CZTI with \swift\ can give spectral parameters similar to that obtained by \fermi/GBM and CZTI can localise this GRB correct to a few degrees.  We also present some results of the measurements of hard X--ray polarisation in this GRB and show that CZTI will be very useful in constraining the emission mechanism in the prompt phase of GRBs.  A detailed description of CZT Imager is given in \citet{Bhalerao2016} and details of onboard performance and spectral fitting methodology are given in \citet{Santosh16} and \citet{Tanmoy16}. For the sake of completeness, however, some salient technical details of CZT Imager are given in the following sections while discussing results like light curves, spectra, localisation, and polarisation.

\section{Observations and data reduction}
\subsection{\swift\ BAT}


The \swift--BAT data were retrieved from HEASARC's data outlet\footnote{HEASARC archive: \url{https://heasarc.gsfc.nasa.gov/cgi-bin/W3Browse/w3browse.pl}}. Lightcurves with 1 s time bin in different energy ranges were made by making use of \sw{HEASOFT-6.17}, \sw{FTOOLS} and the recipe as described in \swift--BAT software guide\footnote{\swift--BAT guide: \url{http://swift.gsfc.nasa.gov/analysis/bat\_swguide\_v6\_3.pdf}}.  We applied gain correction using \sw{bateconvert}, then \sw{batbinevt} was utilised to produce lightcurves after making a detector plane image (dpi), retrieving problematic detectors, removing hot pixels and subtracting the background using \sw{batbinevt} again, \sw{batdetmask}, \sw{bathotpix} and \sw{batmaskwtevt}, respectively. The background subtraction is an advantage with coded aperture masked detectors.

The time integrated \swift--BAT spectrum is obtained in the BAT mission elapsed times (METs) corresponding to the times of our selection for joint time-integrated spectral analysis ($465818107.555$ -- $465818198.115$ BAT MET).  The steps followed to get BAT spectrum are the same as described above to obtain BAT light curves and additional \sw{FTOOLS}  \sw{batupdatephakw} and \sw{batphasyserr} are used for compensating the observed residual in the responses and for making sure that we have the position of the burst in instrument coordinates. We have generated the detector response matrix (DRM) using \sw{batdrmgen}.

\subsection{\fermi}
\fermi--GBM has 12 thallium activated sodium iodide (NaI) detectors and 2 Bismuth Germanate (BGO) detectors, covering energy ranges 8.0 keV -- 1000 keV and 200 keV -- 40 MeV respectively~\citep{Meegan2009}.  The NaI 0, 1, 3 (now onwards referred to as n\#, where \# is the detector number) registered higher fluence than the other NaI detectors as seen in the quick look data from the \fermi\ GRB burst catalog on HEASARC\footnote{\fermi\ GRB catalog: \url{ https://heasarc.gsfc.nasa.gov/W3Browse/fermi/fermigbrst.html}}. Time tagged event data are available for the complete range spanning the T$_{90}$ of this GRB and we make use of this data for both timing and spectral analysis. We used BGO 0 for timing and spectral analysis of the GRB. We choose n0 for making GBM/NaI light curves. The background was fitted in the time intervals $-100$~s to $-10$~s and $100$~s to $300$~s w.r.t. GBM trigger time\footnote{Throughout this paper, time is referred everywhere with reference to \fermi--GBM trigger time, unless otherwise mentioned.} using GBM software \sw{rmfit~4.3.2}\footnote{\url{ http://fermi.gsfc.nasa.gov/ssc/data/analysis/rmfit/}.}.  The light curves were re--binned to $1$~s and obtained in the energy bounds 8--25~keV, 25--50~keV, 50--100~keV and 100--200~keV for n0. For BGO 0 we made a light curve in the 200 -- 500~keV region using again the same tool as n0.

For spectral analysis with the \fermi, we chose the brightest three NaI n0, n1 and n3 and BGO 0 as before. We used {\em rmfit} software to extract the time integrated spectrum from the Time Tagged Event (TTE) files.
The background spectrum was extracted in the time intervals that are specified earlier.
For the time integrated spectral analysis of this GRB, we choose an interval of $-5.5$~s to $85.2$~s with respect to the GBM trigger time. The spectral response for each detector is provided by the instrument team as RSP2 file that contains response for each 2 degree change in the pointing of the \fermi\ spacecraft. 
Our selected time interval was spread over two extensions, therefore we generated a weighted response using \sw{gtburst}\footnote{\fermi\ \sw{gburst} tool: \url{http://fermi.gsfc.nasa.gov/ssc/data/analysis/scitools/gtburst.html}.} tool of \sw{Fermi science tools}.

\subsection{CZTI}
Data from CZTI consist of individual time-tagged photon information for the X--rays registered in the CZT detectors. This information contains identification of the position of interaction (pixel ID), energy of the event, time of registration of the event correct to 20 $\mu$s, information whether there is a simultaneous event from the alpha tagged detector, and information whether there is a simultaneous event in the Veto detector and the energy of the veto event \citep{Bhalerao2016}. The absolute timing has been verified to be better than 200 $\mu$s, while energy information is accurate to 0.5~keV. Data from each quadrant are available separately and independently.

The veto detector provides spectra and lightcurves at 1~s time resolution.
The channel to energy conversion in the Veto detector is done based on the ground calibration and it is estimated that the combined effect of temperature variation, positional dependence etc. can lead to an energy uncertainty of 20\% in the energy range of 50---500~keV.

CZTI data and Veto detector data were reduced using standard \sw{FTOOLS}--compatible pipelines\footnote{CZTI processing pipeline and CALDB files are available under the ``Data and Analysis'' section of the Astrosat Science Support Cell, \url{http://astrosat-ssc.iucaa.in}.} to extract spectra and lightcurves.

\grb\ occurred at an angle of 60\degr.7 from the nominal pointing direction of CZTI, far outside the primary field of view (Figure~\ref{fig:thetaxy}). The coded aperture mask is completely ineffective for localising such sources, and we cannot use the nominal effective area or response files for analysis. The inhomogeneous mass distribution around the CZT detector modules (instrument housing, collimators, etc.; see Figure~\ref{fig:CZTIimage}) results in an energy--dependent transmittivity for every line of sight. Even for a given direction, this transmittivity is different for each CZT pixel.  

We have developed a ray tracing code to estimate the effective area of each pixel in the detector plane for an object at a given location in the sky. The mechanical structure given in Figure~\ref{fig:CZTIimage} is represented by 63 distinct surfaces which are converted into as many cuboids defined by area, thickness, absorbing material, and orientation with respect to the detector surface. For each pixel, the efficiency of transmission through all this material is calculated, along with the detection efficiency of the detectors and geometric projection terms, to give an effective area for a given source direction and energy. The blocking parts of the satellite along these lines of sights have been assumed to be low--$Z$ materials, and have been ignored in this simulation.

For spectral analysis of \grb, \edi{we used the \swift--BAT position to generate effective areas and responses} independently for CZT and the Veto detectors, separately for each quadrant for the latter.  A Gaussian response is generated for CZT with channels numbered from 1 to 512 and Full Width at half maximum (FWHM) of 2.5 keV.  Quadrant wise responses were generated for Veto detector with channels ranging from 0 to 255 and FWHM of 13.04 keV.

\section{Results}

\begin{figure*}[!htbp]
\begin{center}
\includegraphics[angle=0,scale=0.6]{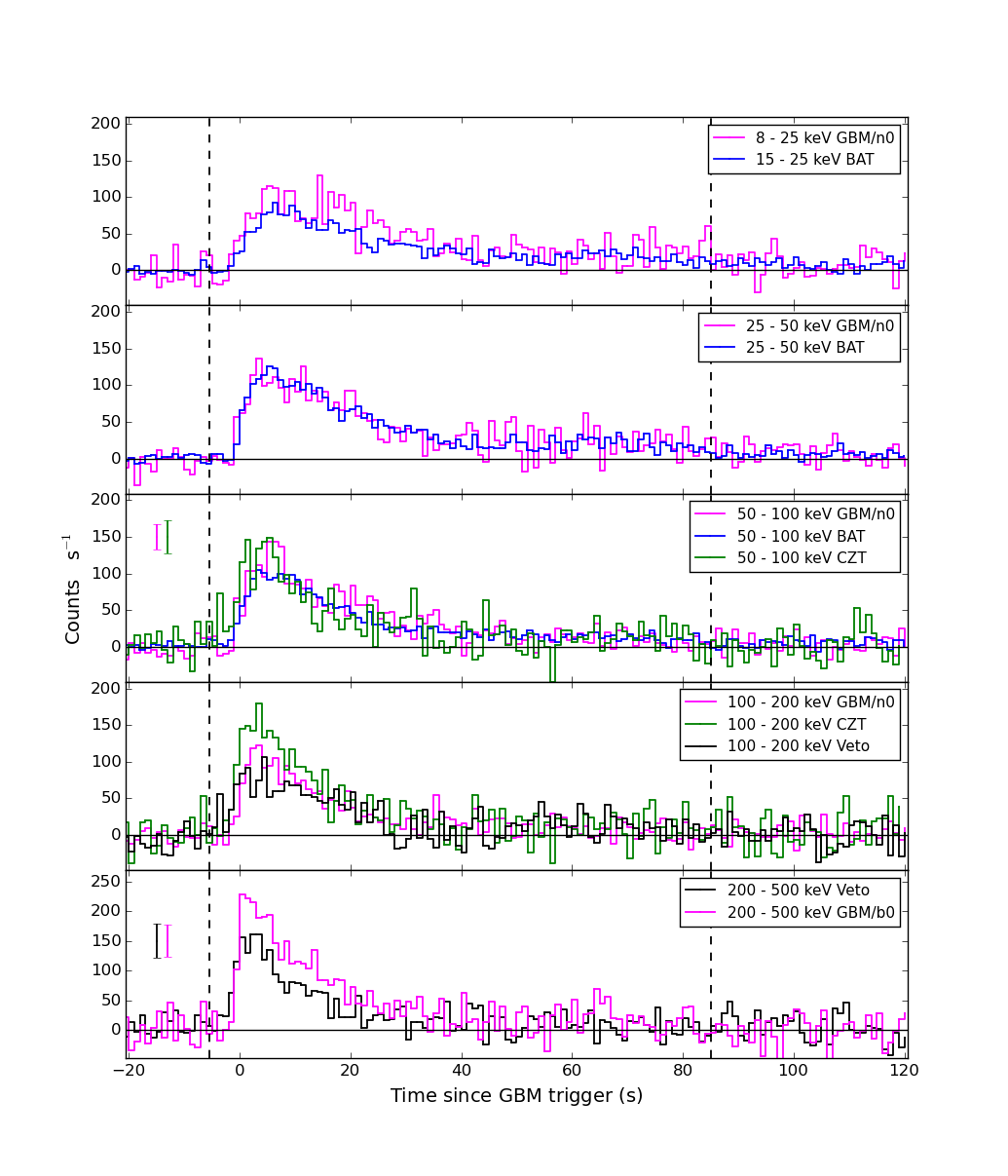}
\caption{Light curves of \grb\  using GBM, BAT, and CZTI data in denoted energy bands. Time is w.r.t. GBM trigger time and the bin size is  1 s. BAT light curves shown here are in counts/sec/illuminated detector and are scaled 1000 times to plot along with GBM light curves. The vertical black dashed lines show the time range used for time integrated spectral analysis. Error bars are not shown, but the typical error bar of CZTI, along with that from GBM, are shown in the third and fifth panels from the top.}
\label{fig:LCs}
\end{center}
\end{figure*}

\subsection{Light Curves}\label{subsec:lc}
The combined light curves of \grb~ using data from \fermi, \swift, and CZTI are shown in Figure \ref{fig:LCs}. The panels are arranged with increasing energy from top to bottom, and data from different detectors are shown in the same panels for similar energy ranges. The observed count rates in BAT, normally given as count~s$^{-1}$ per detector element, are multiplied by 1000 so that they roughly scale to the total observed counts.  The vertical dashed lines indicate the time range chosen for spectral analysis (\S\ref{subsec:spectra}). Errors bars are not shown, however, the typical errors in CZTI data are indicated in the figure with comparative numbers from \fermi.

We fit the light curves with the Norris model \citep{NorrisPULSE}, which describes the temporal profile of GRB pulses with the following equation:
\begin{equation}
I(t) = A \lambda {\rm~exp}{[-\tau_1/(t - t_i) - (t - t_i)/\tau_2]},
\end{equation}
Here, $t$ is time since trigger, $t_i$ is the pulse start time, $\tau_1$ and $\tau_2$ are the scaling times of the pulse rise and pulse decay, $A$ is the pulse amplitude, and the constant $\lambda \equiv \exp{[2 (\tau_1/\tau_2)^{1/2}]}$. The pulse width, $w$,  is derived from the rise and decay time of the pulse as $\tau_2(1+4\sqrt{\tau_1/\tau_2})^{0.5}$, and the asymmetry of the pulse is $\tau_2/w$.  The pulse shape parameters for different energy bands are given in  Table \ref{tab:Norris_model_params} for \fermi\ and CZTI data. It can be seen that the parameters derived from the CZT and Veto detectors  agree with that of the BAT and GBM pulses.  Note that the pulse start time $t_i$ is poorly constrained and when we fix this value for CZTI data to that obtained from GBM, we get comparable values of $\chi^2$.  Further, the measured peak count rates have comparable signal to noise ratio in CZTI and \fermi\ demonstrating that above 100 keV CZTI is as sensitive as \fermi\ for detecting GRBs.  

\begin{table*}
\begin{center}
\caption{A fit to  the pulse profile of \grb\ in tabulated energy bands with the Norris model (\S\ref{subsec:lc}).}
\label{tab:Norris_model_params}
\small
\begin{tabular}{llllllll}
\hline
\hline
Detector & $t_i$ & $\tau1$ & $\tau2$ & $Norm$ & $\chi^2_{red}$ & $w$  & $ \kappa$  \\
 & (s) & (s) & (s) & counts~s$^{-1}$ &  & (s) & \\
\hline\\
\multicolumn{8}{c}{50--100~keV, $E_{mean} = 75$~keV} \\ 
\hline \\
\fermi\ GBM n0   & $~~0.1\pm1.8$  & $0.30\pm0.67$ & $21.0\pm2.1$ &$ 116\pm 10$ & $1.07  $   & $26\pm14$ &$0.82\pm0.52$  \\ 
CZTI       & $-1.4\pm0.6$  & $0.03\pm0.32$ & $17.5\pm2.6$ &$ 93\pm25$ & $0.8$            & $19\pm43$ &$0.92\pm2.2$  \\
~\\
\hline 
\multicolumn{8}{c}{100--200~keV, $E_{mean} = 150$~keV} \\ 
\hline \\
\fermi\ GBM n0    & $-1.43\pm0.26$  & $0.84\pm0.20$ & $15.3\pm1.2$ &$ 102.2\pm5.2$ & $0.99  $  & $21.2\pm2.3$ &$0.72\pm0.13$  \\ 
CZTI         & $-1.43^f$  & $0.03\pm0.02$ & $15.1\pm1.4$ &$ 126.4\pm6.5$ & $0.83  $  & $16.3\pm3.9$ &$0.92\pm0.31$  \\ 
     &  $-4.0\pm0.44$  & $1.29\pm0.76$ & $13.3\pm1.5$ &$ 113.3\pm6.5$ & $0.70  $  & $19.9\pm3.9$ &$0.67\pm0.20$  \\ 
Veto        & $-1.43^f$  & $0.18\pm0.14$ & $18.0\pm2.5$ &$ 85.8\pm6.0$ & $0.58$     & $21.3\pm5.9$ &$0.84\pm0.35$  \\
     &   $-2.8\pm0.9$ & $1.00\pm1.15$ & $16.2\pm2.9$ &$ 81.7\pm7.3$ & $0.57$     & $22.9\pm8$ &$0.7\pm0.38$ \\ 
~\\ \hline 
\multicolumn{8}{c}{200--500~keV, $E_{mean} = 350$~keV} \\ 
\hline \\
\fermi\ GBM b0       & $-1.0\pm0.2$  & $0.00\pm0.17$ & $15.6\pm1.5$ &$ 245\pm929$ & $1.1$ & $15.61$ &$1$  \\ 
Veto      & $-1.14\pm0.6$  & $0.06\pm0.28$ & $12.8\pm1.6$ &$ 158\pm25$ & $0.52$ & $15\pm15$ &$0.88\pm1.04$  \\ 
\hline
\end{tabular}
\end{center}
$^f$:Parameter held fixed in fit.
\end{table*}

\subsection{Spectral Analysis}\label{subsec:spectra}

Many physical and empirical models have been used for GRB spectral analysis, for instance the Band function (\citealt{Band1993}, \citealt{Goldstein2013}, \citealt{Gruber2014}), Band + blackbody \citep{Guiriec2011,Axelsson2012,Burgess2014}, blackbody with a power law 
\citep{Ryde2005,Ryde2009, Page2011, Sparre2012}, double blackbodies with a power law \citep{Basak2015,Iyyani2015} etc. 
Here we examine the time integrated spectrum of \grb\ with a few of the above mentioned models, primarily to emphasise the ability of joint spectral analysis of BAT and CZTI data to produce spectral fit results consistent with those of \fermi\ GBM.   
We make our spectral analysis using four sets of data: a) GBM, b) GBM jointly with CZTI, c) BAT, and  d) BAT jointly with CZTI.

\begin{table}
 \caption{Band model fit parameters for $GRB151006A$.}
 \label{tab:band}
 \begin{center}
 \scriptsize
\begin{tabular}{lr@{ }lr@{ }lr@{ }lr@{ }l} 
Parameter & \multicolumn{2}{c}{GBM} & \multicolumn{2}{c}{GBM+CZTI} & \multicolumn{2}{c}{BAT} & \multicolumn{2}{c}{BAT+CZTI} \\ \hline\hline
 $\alpha$          & -1.1 & $_{-0.1}^{+0.2}$    & -1.08 & $_{-0.13}^{+0.19}$     & -1.2 & $_{-0.1}^{+0.4}$        & -1.22  & $_{-0.18}^{+0.29}$      \\ 
 $\beta$           & -1.8 & $_{-0.1}^{+0.1}$    & -1.75 & $_{-0.1}^{+0.1}$       & -1.71 & $_{-10}^{+0.19}$       &  -1.8  & $_{-0.4}^{+0.26}$       \\ 
 $E_{p}$ (keV)     & 218 &  $_{-78}^{+126}$     & 189.0 & $_{-66.5}^{+87.2}$     & 159 & $_{-82}^{+536}$          & 160.26 & $_{-67.0}^{+214.84}$    \\ 
 $Norm$* & 5.0 &  $_{-1.0}^{+2.0}$    &  5.3 &  $_{-1.0}^{2.27}$       & 3.6 & $_{-1.2}^{+3.9}$         & 4.31   & $_{-1.56}^{3.10}$       \\
 $\chi^2_{red}$    & \multicolumn{2}{c}{0.67}   & \multicolumn{2}{c}{1.10}       & \multicolumn{2}{c}{0.81}       & \multicolumn{2}{c}{1.53}         \\
\hline
\end{tabular}
\end{center}
* Norm is in units of $10^{-3}$~photons~cm$^{-2}$~s$^{-1}$~keV$^{-1}$
\end{table}

\begin{figure*}[bhtp]
\begin{centering}
\includegraphics[angle=270,width=0.49\textwidth]{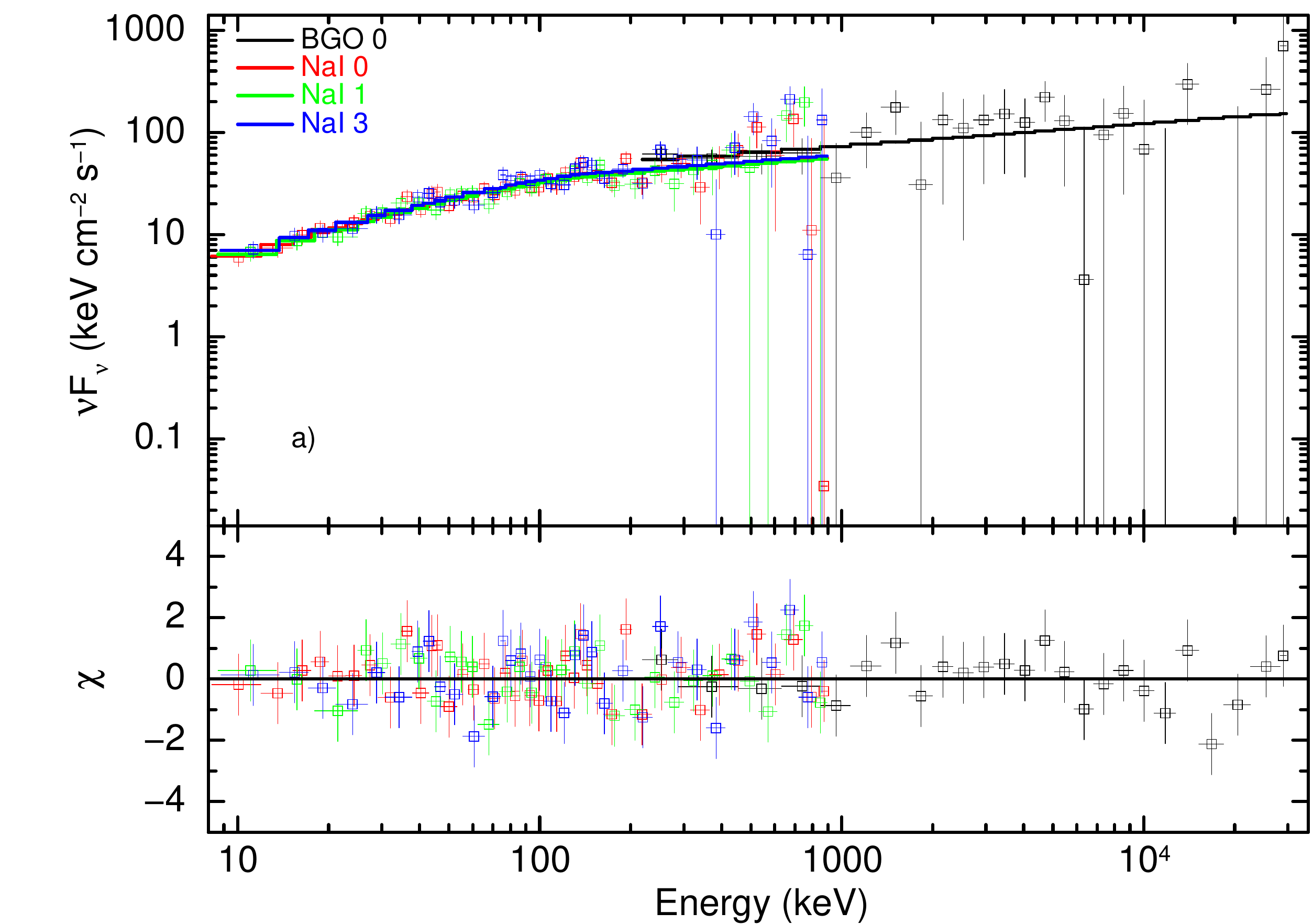}
\includegraphics[angle=270,width=0.49\textwidth]{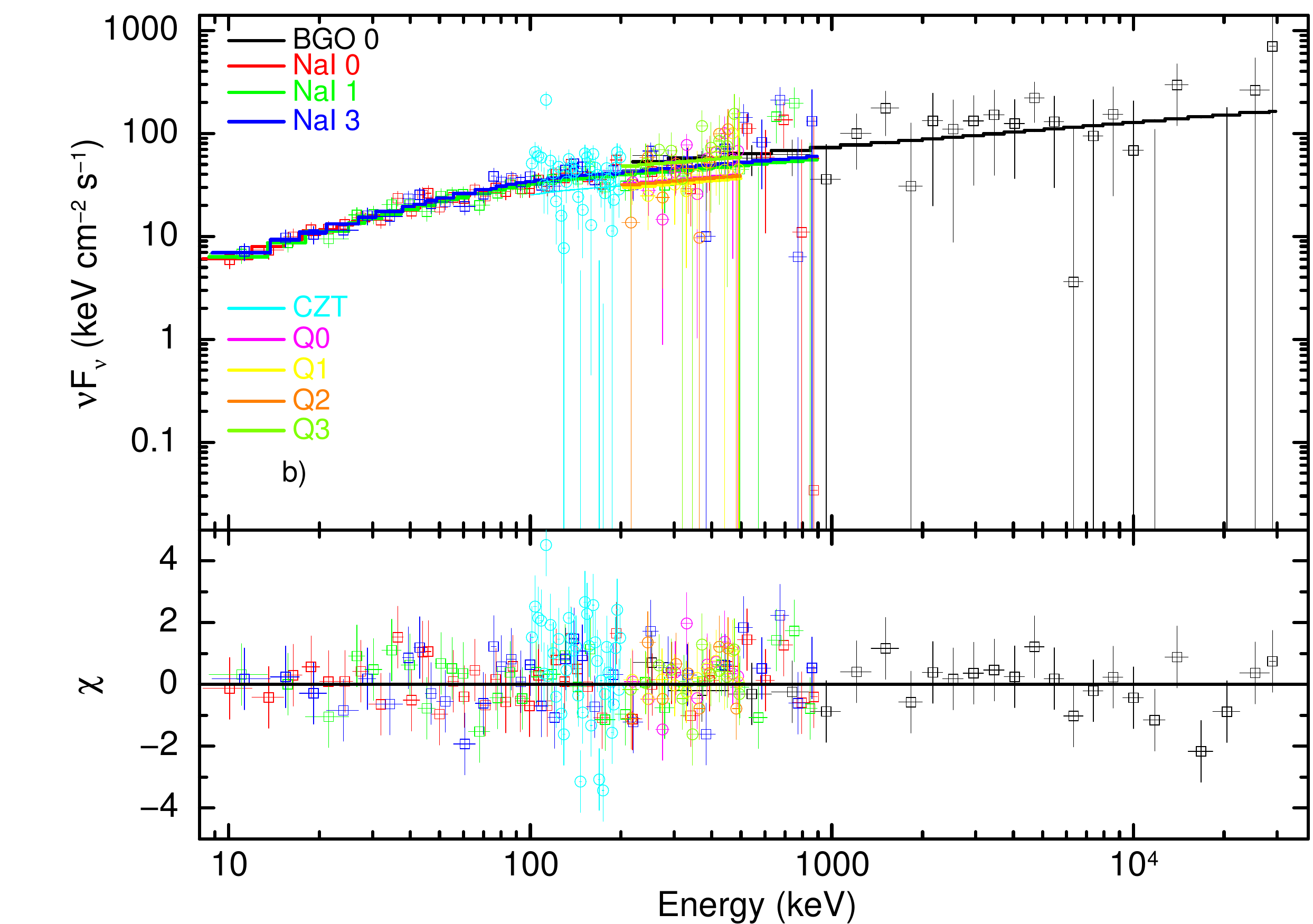}
\includegraphics[angle=270,width=0.49\textwidth]{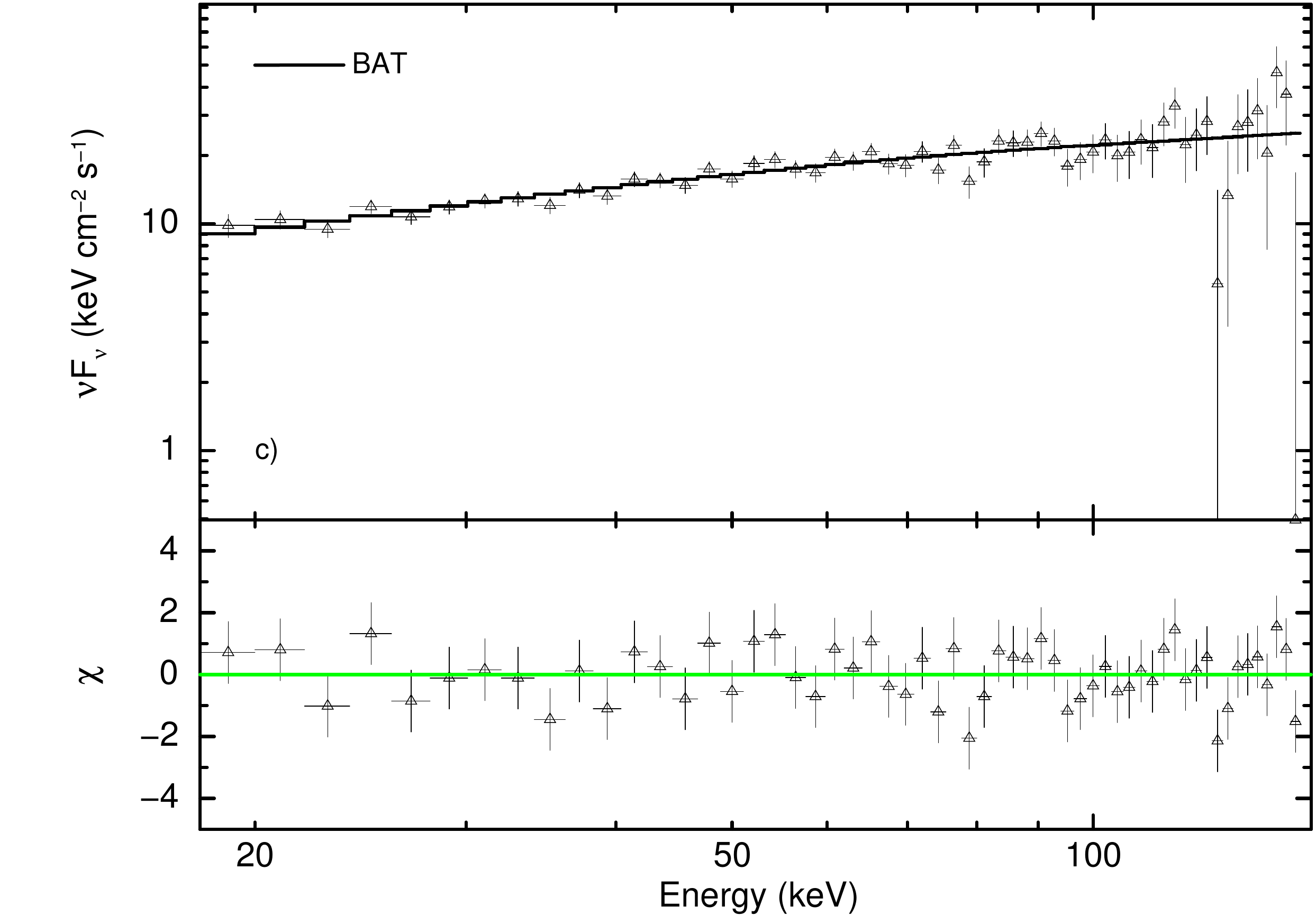} 
$~~~$\includegraphics[angle=270,width=0.49\textwidth]{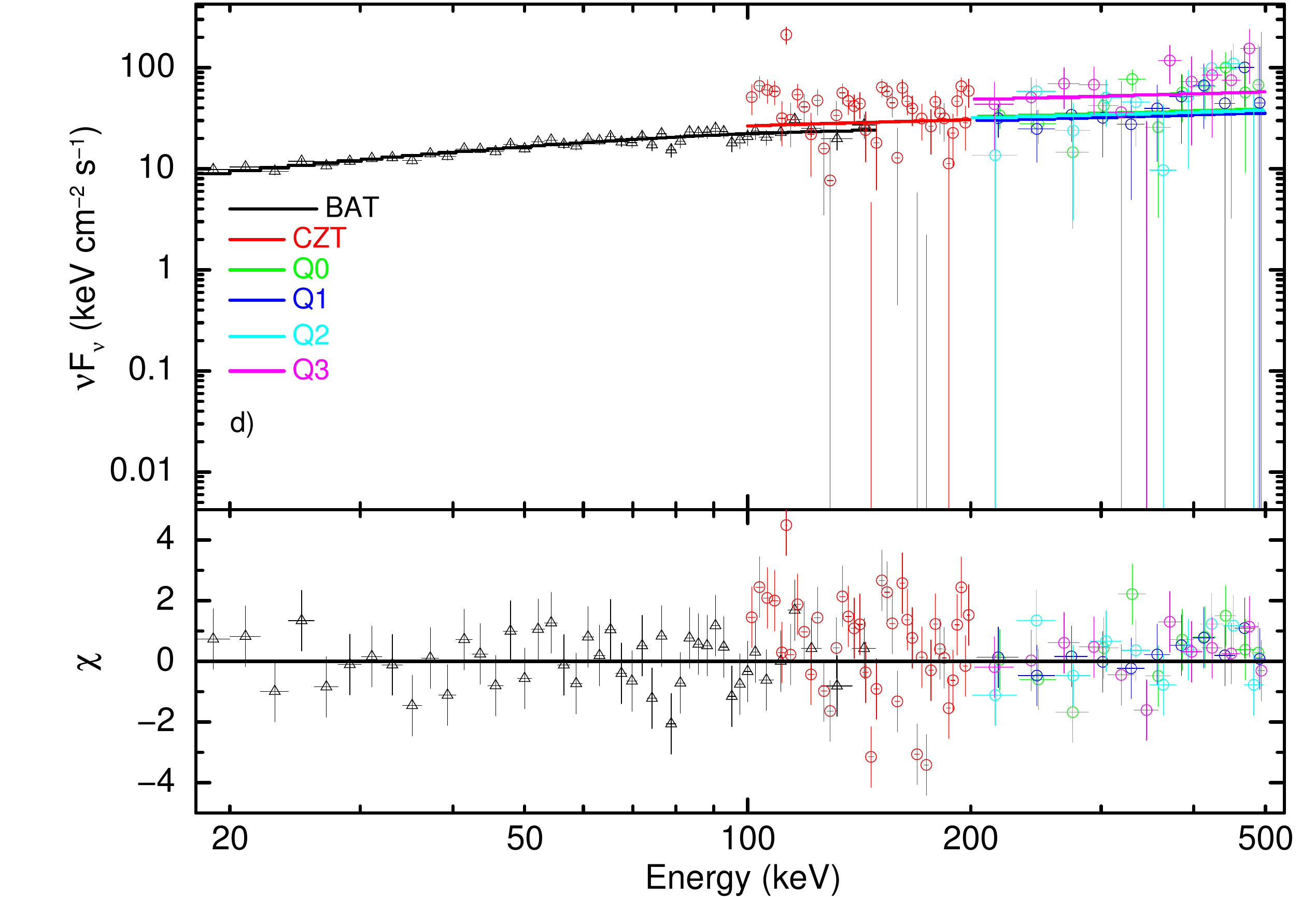} 
\end{centering}
\caption{
The unfolded energy spectrum of \grb\ and the residuals 
\edi{(given in terms of $\chi$, that is  (data - model) scaled to the error in the data)}
are shown based on the spectral fits for the Band model in the time interval -5.5 s to 85.2 s, for the data obtained from a) \fermi\ GBM, b) CZTI combined with \fermi\ GBM, c) \swift\ BAT and d) CZTI combined with \swift\ BAT.
}
\label{fig:GBM_CZTI_Band}
\end{figure*}

To start with, we use the Band model to fit all four sets of data. The best--fit parameters are given in Table~\ref{tab:band}
the data and the results are given in Table 2 and the unfolded spectra along with the residuals, \edi{given in terms of $\chi$, that is  (data - model) scaled to the error in the data}, are shown in Figure \ref{fig:GBM_CZTI_Band}, for all the four sets of data. It can be seen that the high energy index, $\beta$ is not constrained using only the BAT data (due to limited high energy response), but are constrained in all the other data sets.  The obtained values of $\beta$ being greater than $-2$ indicates the presence of high energy emission beyond the currently analysed energy window of the detectors.  
An examination of the residuals shows that CZTI data are consistent with the others, though the use of CZTI data gives a slightly higher value of reduced $\chi^2$.  We also find that the cross normalisation values for CZTI are within 20\% of other detectors. For example, for the BAT data jointly fit with CZTI data, the BAT normalisation with respect to CZT detector is $0.8\pm0.1$, whereas the normalisation of Veto detectors agree with CZT detectors within errors. For GBM jointly fit with CZTI data, the CZT detector normalisation with respect to GBM n0 is  $0.8\pm0.1$.  

The fit values of the parameters obtained for other models (BBPL and 2BBPL), along with the parameters for the Band model, are listed in Table \ref{tab:2bbfits_int} for BAT jointly with CZTI and GBM data to show the effectiveness of using CZTI data with BAT to extend the energy bandwidth.  The models have comparable reduced $\chi^2$s.  The CZTI spectral data have scattered and fluorescent components in $<$ 100 keV region and hence data above 100 keV are considered for spectral fitting.  For the spectral fit with the 2BBPL model, however, the CZTI + BAT data shows a peak at 1 MeV. On careful inspection, the time resolved data showed the presence of multiple thermal components (upto three black bodies) and the time integrated spectra   favoured some of these components depending on the bandwidth of the instruments. A detailed time resolved spectral investigation, including \fermi\-LAT data, would be reported elsewhere.


\begin{table*}[htbp]
 \caption{$Swift/BAT$ + $AstroSat/CZTI$ and $Fermi/GBM$ fit parameters for GRB151006A}
 \label{tab:2bbfits_int}
 \begin{center}
\small
\begin{tabular}{l l l l l l l l l l l}
  Instrument          & Parameters         &  Band               & BBPL                & 2BBPL               &\\ \hline\hline
           &                    \\
BAT + CZTI & $\alpha$/$\Gamma$&$-1.22_{-0.18}^{+0.29}$ & $1.64_{-0.15}^{+0.17}$ & $1.75_{-0.19}^{+0.37}$ &   \\ 
           & $ E_{norm}(keV)/\Gamma_{norm}*$& $~~100.0$     & $4.0_{-1.9}^{+3.7}$ & $6.00453_{-3.2}^{+15.4}$ &  \\ 
           & $E_{p}$/$kT_1(keV)$   &$~~160.26_{-67.0}^{+214.84}$ & $22.1_{-5.4}^{+4.4}$&$23.2_{-4.3}^{+3.6}$&     \\ 
           & $BB_{norm}*$&                           & $0.22_{-0.14}^{+0.19}$&$0.31_{-0.18}^{+0.24}$   &      \\
           &  $\beta$/$kT_2$  &$-1.8_{-0.4}^{+0.26}$ &                    & $ 1098_{-574}^{+\infty}$ &  \\
           & $Band_{norm}$ $10^{-3}$/$BB_{norm}*$ & $~~4.31_{-1.56}^{3.10}$&                &$76.7_{-86.4}^{173.4}$&   \\
           & $\chi^2_{red}$     &$~~1.53$               &$~~1.52 $             & $~~1.52$               &   \\
           &                  &                     &                    &                      & \\
GBM        & $\alpha$/$\Gamma$&$-1.1_{-0.1}^{+0.2}$ & $1.53_{-0.05}^{+0.06}$ & $1.5_{-0.1}^{+0.1}$ & \\ 
           & $ E_{norm}(keV)/\Gamma_{norm}*$& $~~100.0$      & $2.6_{-0.5}^{+0.6}$ & $2.2_{-0.7}^{+1.3}$ &  \\ 
           & $E_{p}$/$kT_1(keV)$   & $~~218_{-78}^{+126}$ & $30.0_{-6.3}^{+7.4}$&$46_{-13}^{+615.4}$&  \\ 
           & $ BB_{norm}*$     &                    & $ 0.3_{-0.1}^{+0.1}$&$0.5_{-0.3}^{+0.2}$&      \\
           &  $\beta$/$kT_2(keV)$  &$-1.8_{-0.2}^{+0.1}$ &                     & $ 12.6_{-3.7}^{+\infty}$&    \\
           &   $Band_{norm}$ $10^{-3}$/$BB_{norm}*$ & $~~5_{-1}^{+2}$&       &$0.15_{-0.1}^{+0.3}$&   \\
           & $\chi^2_{red}$     & $~~0.67$                          & $~~0.7$              & $~~0.69$             &    \\ 
            &    &                      &            &           &    \\
\hline
\\
$*photons~cm^{-2}s^{-1}keV^{-1}$ \\   

\end{tabular}
\end{center}
\end{table*}

\begin{figure*}[!htbp]
\includegraphics[width=\textwidth]{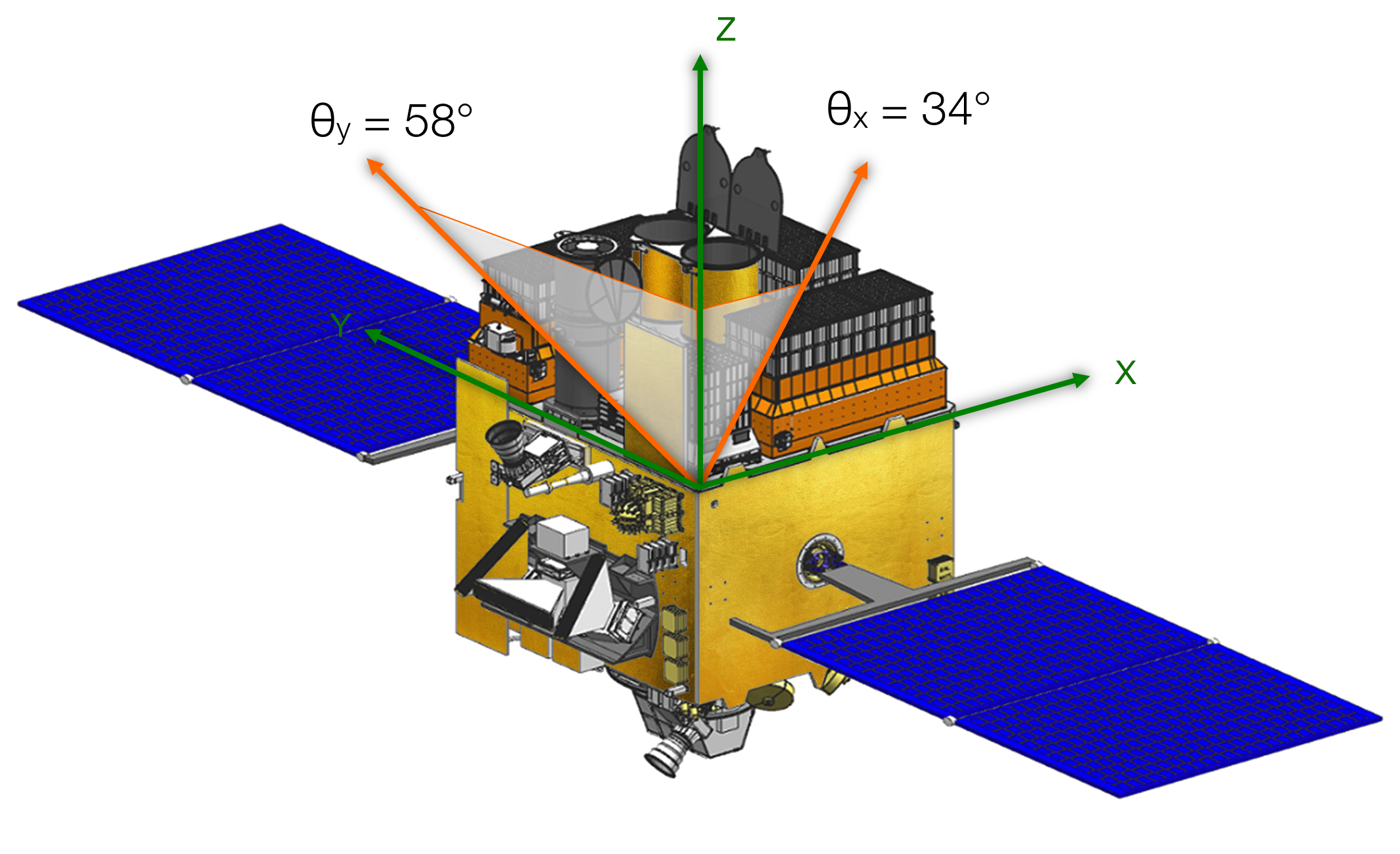}
\caption{A schematic picture of {\em AstroSat} showing the local coordinate definition with respect to the CZT Imager instrument. Localisation is calculated in units of $\theta_x$ ($\theta_y$), angles measured from the $Z$ axis in the $ZX$ ($ZY$) planes. The two components of the incident direction of \grb\ in this coordinate system are indicated.}
\label{fig:thetaxy}
\end{figure*}

\begin{figure}[!htbp]
\includegraphics[angle=0,width=0.45\textwidth]{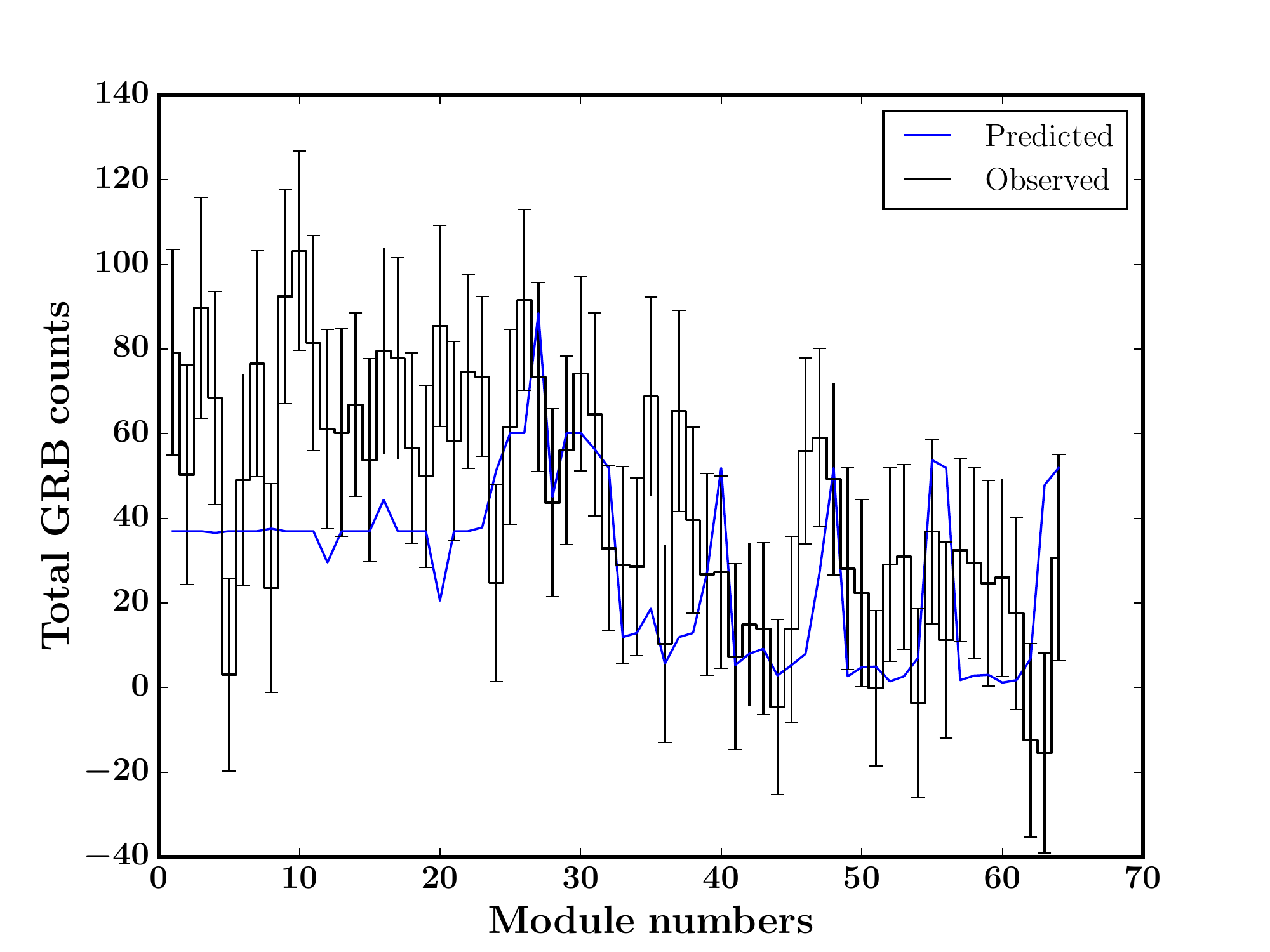}

\includegraphics[angle=0,width=0.45\textwidth]{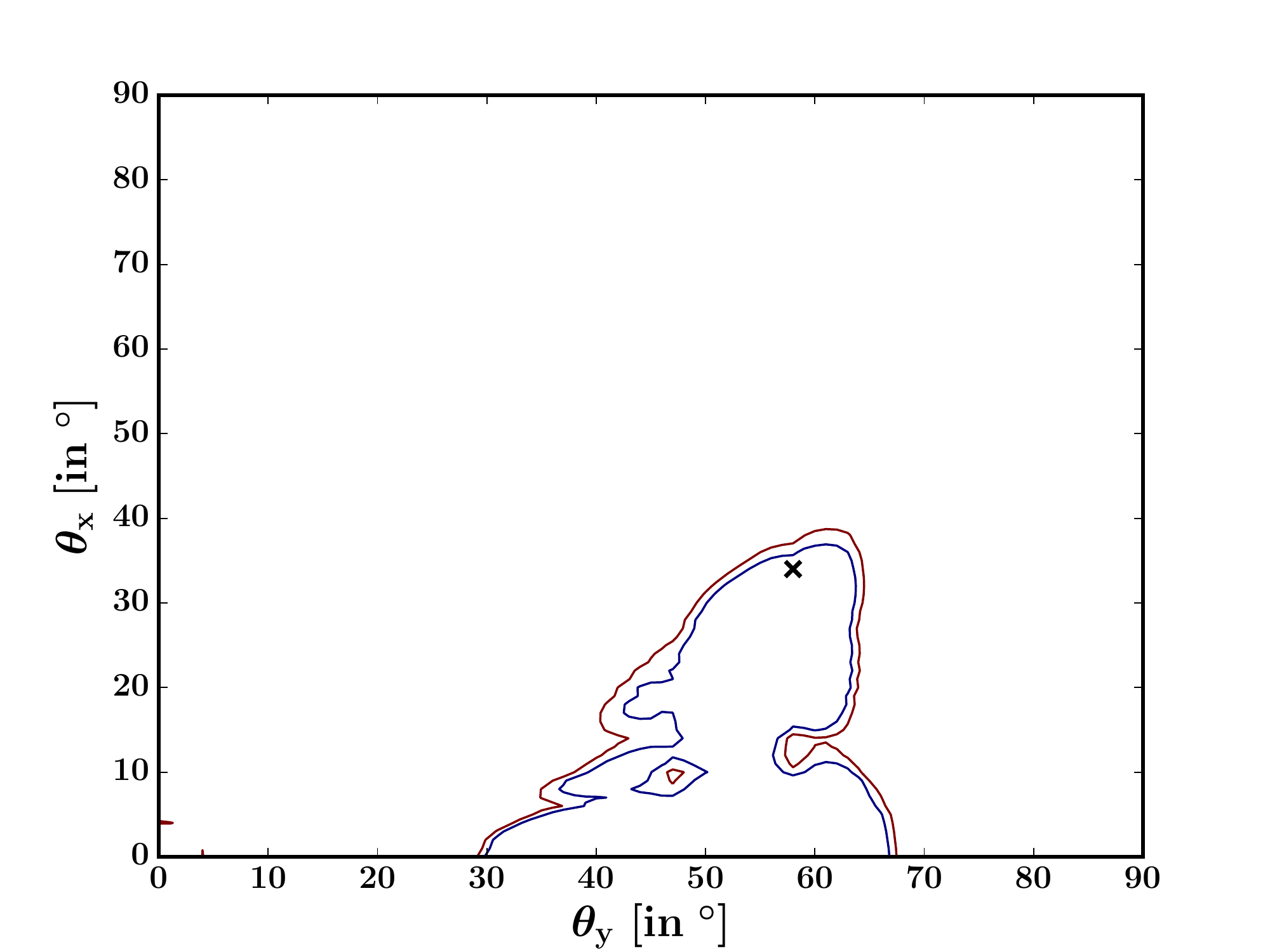}

\includegraphics[angle=0,width=0.45\textwidth]{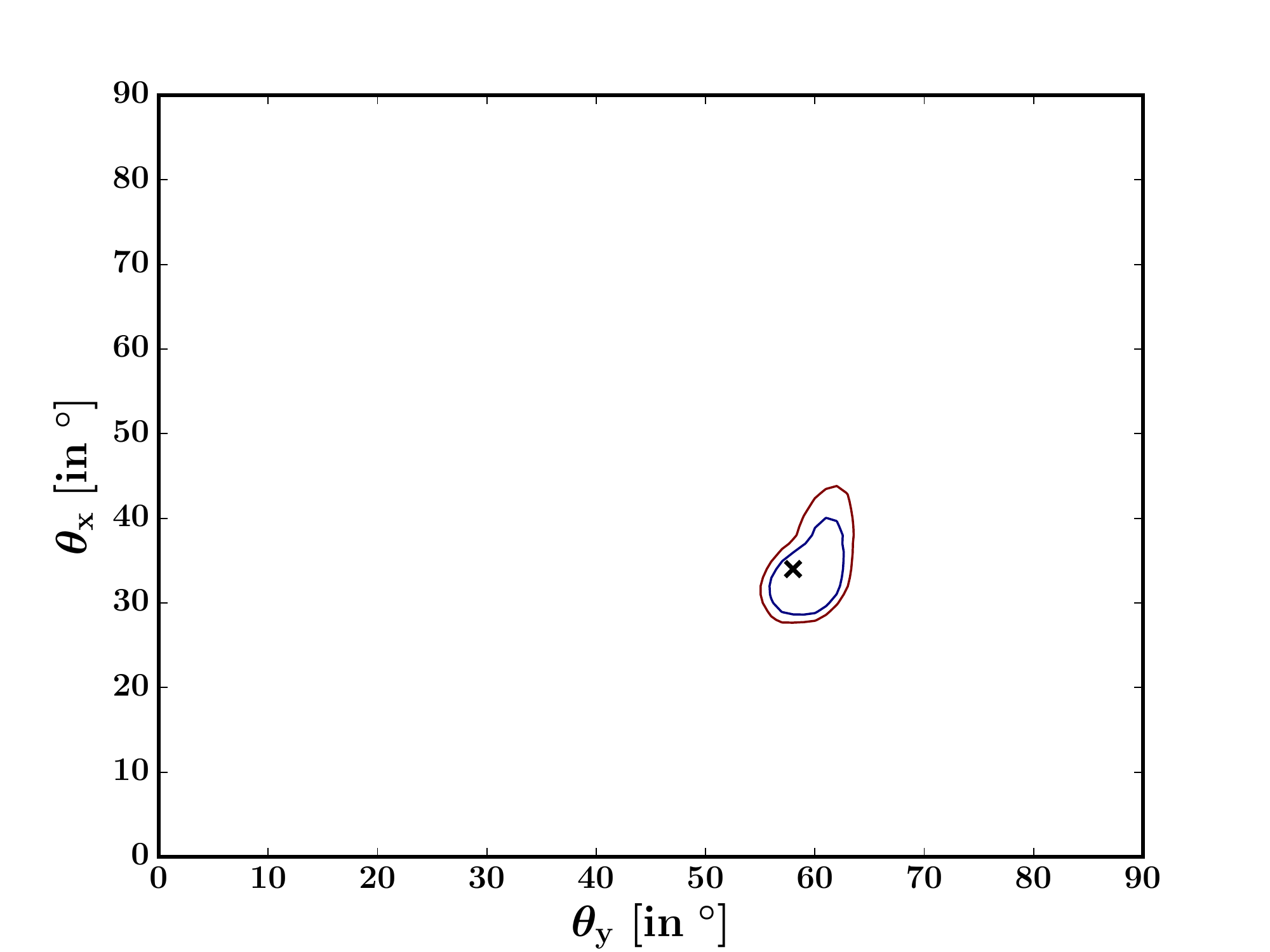}
\caption{ {\em Top:} The observed distribution of counts for the Swift position (see Figure \ref{fig:thetaxy}) in the 64 CZT Modules of $AstroSat$ CZTI instrument for \grb\ (histogram with errors). Each of the modules is blocked by the instrument materials in a different way and this is simulated by a ray tracing model and shown as a histogram. {\em Middle:}  Contour plot of the $\chi^2$ distribution of the observed and predicted counts for predictions based on various incident angles in local coordinates. The position of \grb\ is marked with a cross. The blue and brown $\chi^2$ contours correspond to 90\% and 99\% confidence levels, respectively. {\em Bottom:} The $\chi^2$ contours for a simulated burst with $1$-$\sigma $ errors added by hand. Comparison with the contour obtained from the data indicates the presence of uncharacterised systematic errors in the data.
 }
\label{fig:Localise}
\end{figure}

\subsection{Localisation}


We explore the possibility of using this information to localise transient events like \grb\ occurring outside the FoV of CZTI. Spectra were obtained covering the time duration shown in Figure \ref{fig:LCs} and the background spectra were obtained by adding pre-burst and post-burst intervals of duration  90~s and 200~s respectively. 

We have used the spectral form obtained from the Band model fit  and calculated the pixel wise efficiency for energies in steps of 5~keV. The estimated counts for each of the 64 CZT detector modules, for the Swift position of this GRB (see Figure \ref{fig:thetaxy}) are shown in Figure~\ref{fig:Localise} (top panel) as a continuous line.  Over plotted on this figure are the observed counts. A reasonable agreement between the two are seen.

To test the localisation capabilities of the CZTI, we simulated the instrument response in a grid of $\theta_x, \theta_y$ coordinates (see Figure~\ref{fig:thetaxy}) using the spectrum and fluence of \grb. We then binned the counts by detector, and compared these simulations with the observed values. The $\chi^2$ as a function of local $\theta_x, \theta_y$ coordinates is shown in Figure~\ref{fig:Localise} (middle panel). The actual position of the GRB is marked with a cross. We can see that CZTI could have localised this GRB with an uncertainty of about 10\degr. In order to estimate the contribution of Poisson error to the data, we repeated the exercise by comparing simulated detector--wise counts for \grb\ from the $\theta_x, \theta_y$ grid with simulated counts for the true location, and calculating the $\chi^2$ for each direction (Figure~\ref{fig:Localise}, bottom panel). It is seen that CZTI can localise bright GRBs with an accuracy of a few degrees. The difference between this idealised case and real data may arise from three primary effects: a) non-Poissonian errors in the data due to Cosmic Ray interactions, b) effect of scattering in the detector material, and c) effects of un--modelled absorption in other parts of the spacecraft. A full mass model of the satellite is being prepared and a GEANT-4 simulation for the response of an off-axis GRB is being carried out to understand these systematics. The results of this exercise will be reported elsewhere.\\

%

$~$

\subsection{Polarisation}

The prompt emission from a GRB is expected to be highly polarised owing to the non-thermal origin of the radiation. This is corroborated by the recent findings of high degree of polarisation in the prompt emission of a few GRBs by RHESSI, INTEGRAL and GAP \citep{Coburn2003, Mcglynn2007, Mcglynn2009, Gotz2009, Gotz2013, Yonetoku2011, Yonetoku2012}. 
%
The reported degree of polarisation in the X--ray/gamma-ray band of the prompt emission are in most cases quite high (60--80\%; see review by \citeauthor{CovinoGotz2016} \citeyear{CovinoGotz2016})
 and these are explained in the standard fireball model of GRBs \citep{MeszarosRees1993} as being due to synchrotron emission with an  uniform magnetic field either carried by the outflow \citep{Nakar+2003} or locally produced at the shock \citep{Medvedev2007}. 
 An even higher degree of polarisation can be produced if the primary radiation mechanism is Compton Drag, i.e., inverse Compton emission from relativistically outflowing electrons in the jet \citep{Lazzati+2004}.  A similar mechanism operates in the cannonball model \citep{DadoDar2007} which advocates bulk Comptonisation as the primary source of GRB prompt emission.  On the other hand, evidence of thermal photospheric emission has been reported in many cases of prompt GRB emission \citep{PeerRyde2016}.  This component, while potentially providing seed photons for Compton drag, would by itself contribute little to polarized emission and hence reduce the overall degree of polarisation.

CZTI can help the study of GRB polarisation by measuring X--ray polarisation in the 100--300~keV range. In this energy range, CZTI collimators and support structure are highly transparent. Most of the photons with these energies undergo Compton scattering in pixels of CZTI, and the secondary photon is photo--absorbed in a nearby pixel. The direction of scattering depends on the degree and direction of polarisation of the source, and this effect can be exploited to measure source polarisation~\citep{Tanmoy14, Santosh15}.  An additional advantage of CZTI polarimetry is that the polarisation information can be obtained from the available raw data in standard mode itself without any requirement of changing the hardware configuration.

The photons are expected to be preferentially scattered in the direction perpendicular to the polarisation direction, giving rise to an asymmetry / modulation in an otherwise flat azimuthal angle distribution. Amplitude of the modulation is directly proportional to the polarisation fraction embedded in the incident radiation. True Compton events can be separated from chance two--pixel events by applying three criteria: 1) spatial proximity of pixels, 2) temporal coincidence: events must be recorded within 40~$\mu$s of each other\footnote{CZTI has a time resolution of 20 $\mu$s and we use two clock ticks as the proximity window.}, and 3) the sum and ratio of deposited energies must be consistent with those expected from true Compton events. Selection procedure of the Compton events in CZTI has been discussed in detail in \cite{Tanmoy14}.

\begin{figure}[h!]
\centering
\includegraphics[scale=.4]{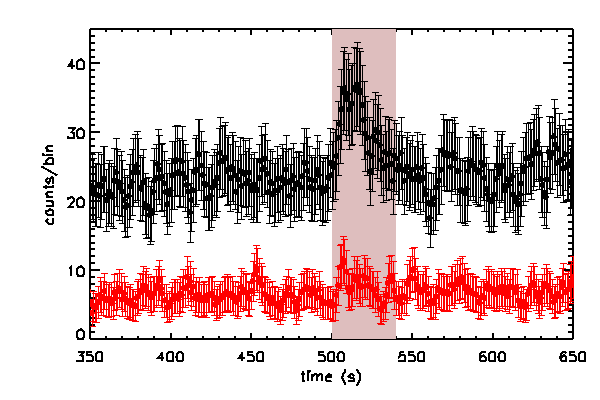}
\caption{The observed rate of double events in CZTI during \grb. The events satisfying the Compton criteria (see text) are shown in black and the red data points are those events not satisfying the Compton criteria. The shaded region in the light curve shows the prompt phase emission of \grb. The Compton events in this region are used for further analysis.}
\label{grbpol_fig1}
\end{figure}

Figure~\ref{grbpol_fig1} shows the one second light curve (with an arbitrary time reference) in Compton events (in black points) for the orbit in which the \grb\ was detected.  Clear detection of the GRB in the Compton events demonstrates the pertinence of the selected Compton events. Furthermore, if we do not  apply the Compton criteria (means selecting only the non-neighboring pixels without putting any Compton energy criteria), the GRB does not show up in the light curve as shown by the red data points.  This gives additional confidence in the selection of Compton events from the raw event mode data.

\begin{figure}[h!]
\centering
\includegraphics[scale=.4]{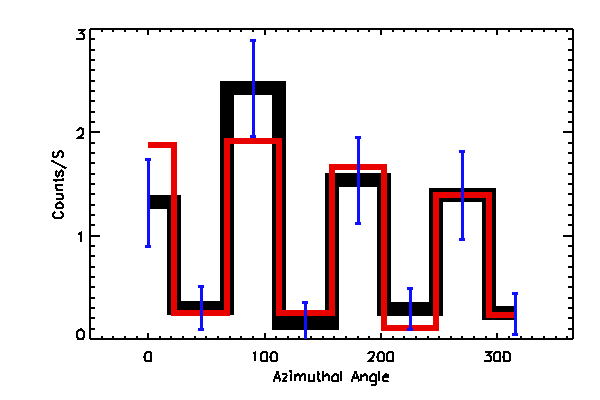}
\includegraphics[scale=.4]{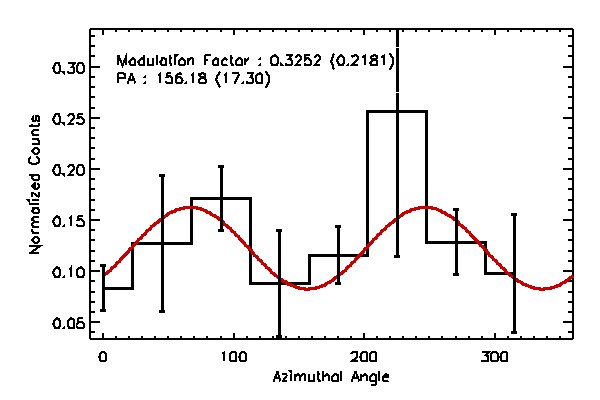}
\caption{{\em Top:} Background subtracted raw eight bin azimuthal angle distribution obtained from the Compton events are shown in black. The error bars shown in blue are the Poisson error on each azimuthal bin for 68$\%$ confidence level. Azimuthal distribution shown in red is the one obtained by simulating with  unpolarised radiation from the same GRB. {\em Bottom:} The geometrically corrected modulation curve for \grb. The red solid line is the $\cos^2\phi$ fit to the modulation curve. The fitted modulation factor is $\sim$0.32 with a detection significance of 1.5$\sigma$. The fitted polarisation angle is $\sim$156$^\circ$ in the CZTI plane. Estimated errors (for 68$\%$ confidence level) on each parameter are given inside the bracket in the text.}
\label{grbpol_fig2}
\end{figure}

The Compton events in the time window of 34 seconds from the onset of the GRB prompt emission (under the shaded region in Figure~\ref{grbpol_fig1}) are analyzed further to obtain their azimuthal angle distribution.  Figure \ref{grbpol_fig2} (top) shows the raw eight bin azimuthal angle distribution for these events (shown in black) after subtracting the background azimuthal distribution from that of the events in the shaded region in Figure \ref{grbpol_fig1}, which contain contribution from both GRB prompt emission and the background.  The background azimuthal distribution is computed from the existing pre-- and post--GRB events. The red bars stand for the simulated azimuthal angle distribution for a 100$\%$ unpolarised beam of spectrum and angle of incidence the same as the GRB.  It is to be noted that the Geant$-4$ simulation for the unpolarised radiation is done with a zeroth order of CZTI mass model. Detailed Geant$-$4 simulation with the complete mass model of CZTI along with the satellite structure and other instruments is currently in progress.  The significant deviation in the azimuthal angle distribution of the observed events from that of the events from an unpolarised beam hints at the presence of a polarisation signature in the prompt emission of \grb, although the statistical significance is low due to the small number of valid Compton events registered from the GRB. The raw azimuthal distribution is then corrected for the geometry of the CZTI pixels as well as for off axis response by normalising the GRB azimuthal distribution with respect to the simulated unpolarised distribution.  The result is shown in Figure~\ref{grbpol_fig2} (bottom). The red solid line in this figure is a $\cos^{2}\phi$ fit to the modulation pattern.  The fitted modulation factor $\mu$ is quite high and it has a value of $\sim$0.32 with a detection significance of 1.5$\sigma$. Such a high modulation factor would imply that the GRB prompt emission is highly polarised.  However, precise polarisation measurement requires the detailed mass model of CZTI along with the satellite and other instruments.  While this is currently being pursued, the result presented above \edi{is a tantalising hint} of a polarisation signature in \grb.

It is to be noted that the Crab nebula has been observed by CZTI for more than 500~ks and a preliminary analysis shows the presence of a statistically significant polarisation signature in the Crab~\citep[in prep]{crabpol}. One of the advantages of GRB polarimetry with CZTI compared to Crab or any other bright persistent X--ray source is the availability of background events immediately before and after the GRB, which is extremely important to quantify the source azimuthal angle distribution. This makes GRB polarimetry with CZTI even more promising.    

Although in case of \grb\ the statistical significance of the obtained modulation is low, the detection of the GRB in the Compton events and thereafter finding a distinct modulation pattern in the azimuthal angle distribution from these Compton events clearly implies that CZTI does have a significant polarisation measurement capability for off axis GRBs, even for those with moderate brightness as \grb.  

\begin{figure}[t!]
\centering
\includegraphics[scale=.4]{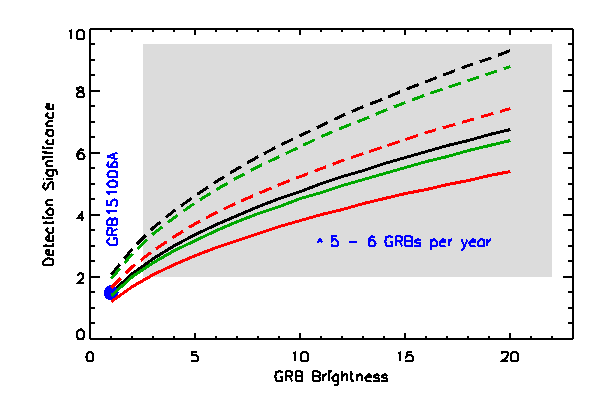}
\caption{Expected polarisation detection significance for GRBs with brightness \edi{(fluence)} in the units of  the brightness level of GRB151006A assuming spectra similar to that of GRB151006A and different polarisation fractions (solid lines: 50$\%$ polarisation, dashed lines: 70$\%$ polarisation) and off axis angles of detection (black: 40$^\circ$, \edi{green}: 60$^\circ$, red: 20$^\circ$). The filled green circle stands for the detection of \grb.}
\label{grbpol_fig3}
\end{figure}

It is to be noted that this is the first time that polarisation signature in a GRB of such a moderate brightness has been reported.  Other GRBs for which polarisation measurements have been reported so far are at least 5--10 times brighter than \grb. Therefore, brighter GRBs detected by the CZTI would certainly yield a considerably higher significance in polarisation detection. Figure~\ref{grbpol_fig3} shows the expected detection significance for brighter GRBs,
with different polarisation fractions and angles of incidence.  Different colors stand for different off axis angles and we see that the polarisation sensitivity of the CZTI is the highest at off axis angles close to 45$^\circ$, due to the higher effective area of the instrument at these angles. The solid and the dashed lines are obtained for polarisation fractions of 0.5 and 0.7 respectively.  For GRBs with fluence three or more times that of GRB151006A, polarisation may be estimated with a significance of more than 2.5$\sigma$. The CZTI is expected to detect polarisation of 5$-$6 such GRBs per year. Statistical analysis of all these polarisation measurements from CZTI along with other upcoming GRB polarimeters will be extremely useful in constraining the existing models of GRB prompt emission mechanism \citep{Toma2008}.

\section{Discussion and Conclusions}

The CZTI detection of \grb, far off--axis, on the very first day of operation demonstrates the capabilities of CZTI as a wide--angle GRB monitor. As the instrument had just been switched on, some of the parameters of the instruments were not finalised and tuned. For example, the low energy threshold was quite high (close to 30 keV instead of the design goal of 10--15~keV). Also, the Veto detector was not operating in anti-coincidence mode as the relevant timing parameters were not yet set. Nevertheless, observations of \grb\ has demonstrated several useful features of the instrument and CZT Imager promises to be a good all sky monitor above 100~keV for transient events. The total effective area of CZT Imager (Figure~\ref{fig:CZTIimage}) is comparable to that of \fermi. The fact that most of the satellite materials are transparent to X--rays above 100~keV makes the field of view of CZT Imager close to 3$\pi$ steradians at these energies. Utilisation of the observed data, however, requires a good description of the transparency of the satellite material and currently we are working on a complete satellite mass model to firm up the estimate of full sky effective area as a function of viewing angle and photon energy.

Since the CZT Imager has a large area ($\sim$980 cm$^2$) detector with good position accuracy, the material distribution of the instrument and the satellite itself can be used as a coder to infer the incident direction of the transient events. Preliminary studies for this GRB has demonstrated that it should be possible to localise GRBs correct to a few degrees. A detailed investigation of several GRBs is being undertaken to better quantify the localisation accuracy. 

The most exciting feature of this new instrument is its capability to measure the polarisation signals above 100 keV \citep{Tanmoy14, Santosh15}.  One good feature of the CZT Imager is the continuous availability of the time-tagged data so that no additional modes are required to be activated to measure the polarisation signature. As mentioned earlier, this is the first time that \edi{hints of a} polarisation signal are reported for a GRB of fluence less than 2 $\times$ 10$^{-5}{\rm~ergs~cm}^{-2}$ and hence for bright GRBs the CZT Imager will provide polarisation information with a vastly superior significance.  
Accurate estimate of the degree of polarisation, measurement of time evolution of polarisation properties and their relation to the spectral evolution have the potential to clearly distinguish between the various suggested models of GRB prompt emission mechanism.  With its good spectral sensitivity, and the capability to detect hard X--ray polarisation, the CZT Imager promises to make a significant contribution to this investigation.  

As an all-sky hard X--ray monitor CZTI has sensitivity comparable to \fermi\-GBM and hence has the ability to detect possible hard X--ray transients associated with gravitational wave events \citep{LIGO2016, LIGOGBM2016}.

In summary, the CZT Imager aboard {\em AstroSat} is a new addition to the suite of GRB instruments 
with an exciting new combination of capabilites such as spectroscopy, polarimetry and localisation. A 
detailed analysis of data from several GRBs is currently under way to fully characterize and refine these 
various features.

\acknowledgements  
This publication uses the data from the AstroSat mission of the Indian Space Research Organisation (ISRO), archived at the Indian Space Science Data Centre (ISSDC).
CZT-Imager is built by a consortium of Institutes across India including Tata Institute of Fundamental Research, Mumbai,  Vikram Sarabhai Space Centre, Thiruvananthapuram, ISRO  Satellite Centre, Bengaluru,  Inter University Centre for Astronomy and Astrophysics, Pune, Physical Research Laboratory, Ahmedabad, Space Application Centre, Ahmedabad: contributions from the vast technical 
team from all these institutes are gratefully acknowledged. We are thankful for the helpful discussions with K. L. Page, E. Troja and C. Markwardt for $Swifthelp$ facility.
This research also has made use of data obtained through the High Energy Astrophysics Science Archive Research Center Online Service, provided by the NASA/Goddard Space Flight Center. RB is a stipendiary of START program of the Polish Science Foundation (2016) and supported by Polish NCN grants 2013/08/A/ST9/00795, 2012/04/M/ST9/00780, 2013/10/M/ST9/00729, and 2015/18/A/ST9/00746.
\newpage
\bibliographystyle{apj} 
\bibliography{grb}

\end{document}